# Towards a More Well-Founded Cosmology

Hartmut Traunmüller

Stockholm University, Stockholm, Sweden

e-mail: hartmut@ling.su.se

## Abstract

First, this paper broaches the definition of science and the epistemic yield of tenets and approaches: phenomenological (descriptive only), well-founded (solid first principles, conducive to deep understanding), provisional (falsifiable if universal, verifiable if existential), and imaginary (fictitious entities or processes, conducive to empirically unsupported beliefs). The Big-Bang pardigm and the ΛCDM "concordance model" involve such beliefs: the emanation of the universe out of a non-physical stage, cosmic inflation (invented ad hoc), Λ (fictitious energy), and exotic dark matter. They fail in the confidence check that is required in empirical science. They also face a problem in delimiting what expands from what does not. In the more well-founded cosmology that emerges, energy is conserved, the universe is persistent (not transient) and the 'perfect cosmological principle' holds. Waves and other perturbations that propagate at $c$ (the escape velocity from the universe) expand exponentially with distance. This dilatation results from gravitation. The cosmic web of galaxies does not expand. Potential $\Phi$ varies as $-H/(cz)$ instead of $-1/r$. Inertial forces arise from gravitational interaction with the rest of the universe (not with space). They are increased where the universe appears blueshifted and decreased more than proportionately at very low accelerations. A cut-off acceleration $a_0 = 0.168\ cH$ is deduced. This explains the successful description of galaxy rotation curves by MoND. A fully elaborated physical theory is still pending. The recycling of energy via a cosmic ocean filled with photons (the CMB), neutrinos and gravitons, and wider implications for science, are briefly discussed.

## 1. Introduction

Empirical science involves acquiring knowledge with an aim to organize, explain and understand phenomena. This knowledge is in part 'existential' and in part 'universal' (such as physical laws). The latter type of knowledge can be conceived of as *a set of empirically testable universal claims that have not yet been convincingly falsified* and so remain *tenable.* In addition to this, which is most prominent in Popper's (1935) philosophy of science, science makes also *existential claims*, of which Pauli's prediction of the existence of neutrinos (Mößbauer, 1998) is a non-trivial example. Existential statements can only be *verified* rather than falsified empirically, unless they claim something to exist at a specific place and time. In this conception of science, it is in neither case necessary for the postulates and hypotheses, which give rise to the claims, to be understood. It suffices for them to be tenable given the empirical evidence. However, it can be argued that the ultimate aim of basic research is to extend *the body of empirical knowledge that can be rationally explained ab initio*, i.e., without reliance on any assumption that is not understood. Assumptions or hypotheses that are not understood still have a function in science at a less advanced stage of development, but they are bound to remain tentative and provisional until they are shown to be either redundant (= predictable within a wider frame) or untenable.

In order to really *understand* phenomena and the relations between these, we need theories that rest on a foundation of solid knowledge. This may involve other well-founded, more fundamental theories. Ultimately, well-founded theories are based solely on *definitions* and *first principles* of the kind that cannot easily be rejected using the Cartesian method of doubt. These are principles that are accepted even outside the frame of the particular theory. Some are indispensable for there to be a theory at all. In the present paper, the notion of 'first principle' is always to be understood in this narrow sense. An 'axiom' does not necessarily qualify as a first principle in this sense.

Indispensable axioms whose validity is independent of nature lie at the foundation of the formal sciences. These give us the rules of logic, algebra and geometry, which then can be taken as first principles in all sciences. It may not always be clear what can be taken as a first principle, but many theories build on a postulate that can easily be called into question. In such cases, it is clear beyond any doubt that the postulate does not qualify as a first principle in our sense. A theory that depends on it cannot be more than a speculative, conditional and provisional one, even if its predictions are compatible with all available empirical evidence, no matter how accurately. It will remain 'just a theory' even if 'corroborated' by evidence. While many theories are of this kind, there are also more well-founded *ab initio* approaches.

Physical *ab initio* approaches have been pursued in chemistry (e.g. ab initio quantum chemistry, ab initio molecular dynamics) perhaps more often than in physics itself. In physics, there is a strong tradition of attempting to reconcile empirical knowledge with a few traditional standard paradigms that may fall short of satisfying the mentioned criteria of well-foundedness. It is well known that inferior paradigms and standards can persist because of the legacy they have built up, like the QWERTY layout in typewriters (David, 1985). Such "path dependence" is also prominent in the history, teaching and practice of science. This had, in effect, already been noticed by Kuhn (1962) in his study of scientific practice, but the undesirable "lock-in effects" of path dependence have not yet found the attention they require there. These have been mainly discussed in the field of economics, and the few papers on path dependence in epistemology also originated there (Jolink & Vromen, 2001, Peacock, 2009).

The history of science shows us that questionable assumptions on which previously established theories had been based tend to be retained not only as long as they remain compatible with the empirical evidence but as long as they can be made compatible with it by ad hoc means. Standard cosmology is a prominent case in point, and it had a precursor already in Newton's questionable treatment of inertia as an effect of space (not of the matter in it), which Einstein retained and extended in General Relativity (GR).

In current standard cosmology, the Big Bang (BB) paradigm is taken for granted. Due to its free parameters and liberal allowance for evolution, it is flexible, but it happened that new or previously neglected evidence was found to be incompatible with it nevertheless. In such cases, a theory stands falsified until a convincing explanation of the discrepancy is presented. Although this is clear enough, it is not very rare in scientific practice that falsifications are brushed aside by advancing excuses in the form of ad hoc assumptions and constructs, also purely imaginary ones, which can only be believed in. Such adherence to traditional paradigms is characteristic of what Kuhn (1962) called "normal science" as opposed to "revolutionary science" and Lakatos (1976) a "research programme". It is advantageous for those who aim for or depend on positive judgments by teachers, referees, editors and grant providers, and for extensive collaboration. However, approaches that require ‘credence’ in ad hoc assumptions can, in the long run, hardly be claimed to remain within the bounds of ‘science’ at all. They are symptomatic of a degeneration of the science into a fossilized system of unquestioned doctrines.

We shall take a look at the epistemological status of the assumptions that have led to the standard model of BB cosmology, the ΛCDM concordance model, and contrast this model with the implications of alternatives in which ad hoc solutions are avoided and the most deeply rooted one of the questionable physical tenets, the association of inertia with space, is dropped, while conservation of energy is taken as a first principle and the “perfect cosmological principle” (PCP) as a generalizing assumption. The latter implies that the universe is persistent instead of transient. It will be shown that the astronomical evidence that requires excuses in order to maintain the BB paradigm appears compatible with a persistent universe.

## 2. Method: Confidence Check

The common definition of “empirical science” as ‘the pursuit of knowledge about nature’ is not accurate enough for our purposes. In addition to “knowledge”, traditionally defined as ‘justified true belief’, we must allow for beliefs or, more objectively, for “statements” that have only been shown to be ‘tenable’ rather than ‘true’, while accidental truths of the type described by Gettier (1963) and unsystematic statements have to be excluded. This is achieved by substituting “reliable systematic statements” for “knowledge”.

*Empirical science is the pursuit of reliable systematic statements about nature.*

This definition requires taking the confidence that premises deserve, and on which depends the empirical reliability of conclusions, into account. It dismisses approaches that fail in a confidence check or merely reject such a check. It also dismisses untestable hypotheses, the reliability of which cannot be checked. Although the definition implies that science strives for ultimate reliability, it defines science as a pursuit, and tentative premises, hypotheses and statements have a place in this pursuit as long as their reliability cannot be denied. Development and use of improved tools and methods is an integral part of the scientific pursuit that often contributes to its progress; but we are here not concerned with applied science, i.e., with the art of using science for the solution of practical problems.

Among scientific approaches to natural phenomena one can distinguish between inductive, phenomenological ones, which are founded on observations, and deductive ones, which are founded on theoretical premises. There is often interplay between these, e.g., an inductive approach may suggest a hypothesis that is subsequently used in a deductive approach. Definitions are essential in both types of approach. What distinguishes the approaches is the kind of conclusions the respective premises allow to be drawn with confidence and the resulting epistemic yield. The third type of reasoning, abductive inference, involves both induction and deduction.

In purely phenomenological approaches (type 1 in Table 1), regularities among observations (occasional evidence) are searched and described without offering an explanation. They yield organized existential knowledge, empirical relationships, and superficial or probabilistic understanding. Phenomenological models make use of formalisms and free parameters. A well-known example is present in Kepler’s laws of planetary motion. Exploratory data analysis is an archetypal method. While occasional evidence can provide conclusive support ($C$ = 1) for an existential statement, approaches of this type can only provide suggestive and statistical support for universal statements.

**Table 1.** Epistemologically different types of scientific approaches and tenets (1: inductive, 2: deductive), the confidence $C$ these impart (a multiplicative variable), their type of adequacy and their epistemic yield.

| | Premises | Foundation | Confidence | Adequacy | Epistemic Yield |
|---|---|---|---|---|---|
| 1 | **Occasional evidence** | definitions + observations | $0 < C \leq 1$ | descriptive | superficial and/or probabilistic understanding |
| 2a | **Well-founded tenets** | definitions + first principles | $C = 1$ | descriptive + explanatory | deep understanding (ab initio) |
| 2b | **Provisional tenets** | " + tentative assumptions | $0 < C < 1$ | descriptive + tentatively explanatory | superficial and uncertain deeper understanding |
| 2c | **Fictitious tenets** | " + fictitious assumptions | $C = 0$ | formal | empirically unsupported belief |

Theoretical, deductive approaches offer, in addition, an explanation of observations. They can provide conclusive support for universal statements, but only to the extent to which we can be confident in their premises. Therefore, it is necessary to distinguish at least three epistemologically different types according to the roots of the tenets they profess (type 2a, 2b and 2c in Table 1).

2a) First principles. In cases in which these are sufficient, they lead to well-founded theories and predictions and to explanations that can be understood ab initio. Approaches that are founded on definitions and first principles alone embody the deepest understanding of phenomena. However, first principles can be invoked in all deductive approaches and even in otherwise empirically founded ones.

2b) Tentative assumptions, also called “postulates”, that in some way appear reasonable but remain subject to doubt since they are not rooted outside the theory in question and can never be proven within it. These lead to provisional (conditional) theories and to explanations that hold to the extent to which the assumptions hold. This is the type of approach envisioned in the “hypothetico-deductive method” of science, which is prevalent in theoretical physics inside and outside the mainstream, but which fails to distinguish between the types 2a, 2b and 2c, whose premises differ grossly in the confidence they deserve.

2c) Assumptions that, in addition to not being rooted outside the theory in question, also lack independent empirical support. Any reasoning based on these remains within the domain of imagination. Such assumptions are ‘fictitious’[1] and lead to epistemically void beliefs. Modern theoretical physics offers a range of “fairy tale physics” (Baggott, 2013) in which fictitious assumptions are either primary, as in string theory, or secondary, as in the "dark sector" of BB cosmology, discussed in Section 3.

[1] By “fictitious”, we mean ‘merely existing in theory, not in reality’. In contrast, so called “fictitious forces” are never fictitious in this common sense, but rather in the opposite sense, which reflects a theory-centered world view that is characteristic of theoretical physics.

The values listed in Table 1 under "Confidence" express the confidence we can have in the tenets and the explanations these suggest. They depend on how well the tenets are rooted in what is already understood. We can be fully confident if the tenets are well-founded (type 2a). If they really are, our confidence remains undiminished even when confronted with discrepant empirical data. If, on the other extreme, an entity or process is fictitious within the frame of existing knowledge (type 2c), the conficence it deserves, its explanatory power and its epistemic value are no larger than zero, $C = 0$, exactly. This holds even if the approach leads to predictions that are compatible with the evidence. The provisional approaches (type 2b) lie between the extremes 2a and 2c ($0 < C < 1$). In these cases, a numerical rating of confidence that would be generally valid is not obvious, except at the level of rank order. It is, e.g., justified to attach more confidence to a reasoning based on a generalizing assumption that has not been falsified than to an alternative that can be said to involve the same assumption under a restrictive condition that needs to be specified. The latter is equivalent to having two assumptions instead of just one, and the higher confidence in an approach that needs fewer assumptions reflects the principle of parsimony (Ockham's razor), which applies here.

Sufficiently, even fully reliable predictions of entities that have never been observed are not precluded in this scheme. Such entities are not necessarily fictitious within the frame of existing knowledge. In order for us to be confident at $C > 0$ into their real existence, it is only required that $C > 0$ for each of the tenets on which the prediction is based.

It is well known that empirical falsifications are not firmly conclusive. This is part of the Duhem-Quine problem. Falsifications are only valid within the frame of the knowledge we have. A statement that stood falsified may even become tenable again in the light of new knowledge. Strictly speaking, universal statements can only be claimed "to be tenable" or "to stand falsified", unless it follows from definitions and logic alone that they are "true" or "false". The classification of a tenet as fictitious ($C = 0$) might also change in the light of new knowledge, but as long as we lack this knowledge, our confidence in it must remain at zero if we wish to remain within science.

When confronted with discrepant evidence, the descriptive adequacy of a theory can often be saved by introducing an ad hoc parameter. However, such a parameter has no explanatory power. Worse yet, it invites circular reasoning, and if it represents a fictitious entity, the approach turns into one of type 2c. This yields just an epistemically unsupported belief ($C = 0$), e.g. in dark energy. It promotes 'credence' – not 'science'.

Some first principles with $C = 1$ can be derived logically on the basis of more widely valid principles or well-founded theories, but ultimately there remains a basic physical principle that can neither be verified logically nor empirically with full certainty, despite its wide range of empirically proven tenability. Its classification as a first principle is, instead, due to its being indispensable for there to be any 'law of nature' and any explanatory science at all. It says that *the same physical laws are valid everywhere in space, direction and time*. This universality principle expresses a precondition for physics.

There are several conservation laws that can be derived via Noether's theorem from the homogeneity of space-time that is implied in the universality principle: conservation of energy follows from the homogeneity of time, conservation of linear momentum from that of space, and conservation of angular momentum from the isotropy of space. While these may be shown to follow from one general symmetry principle, theories do not gain in confidence if the number of first principles they invoke (all with $C = 1$) is minimized. They gain in confidence if the number of tentative assumptions (all with $C < 1$) is minimized, provided that they do not involve any fictitious assumption (with $C = 0$).

While homogeneity and isotropy of space-time belong to the set of first principles, this is not equally clear for the PCP. The PCP states that the distribution of matter in the universe is isotropic and homogeneous in space and in time.[2] Bondi and Gold (1948) considered it a first principle that gives us a reason for assuming that the same physical laws are valid everywhere, but it appears that the matter distribution in the universe is more in need of an explanation than the universality principle. The PCP is at least a clear case of a generalizing assumption. The (imperfect) cosmological principle adopted in the BB paradigm is not. It exempts the temporal dimension of space-time, which is possible in an absolute system of reference but not generally in Minkowski space-time.

The foundational elements of theories can be listed in the following order:

- definitions
- well-founded first principles
- generalizing assumptions
- more specific testable assumptions
- assumptions involving fictitious entities or processes.

In order to obtain a radically more well-founded theory, it is necessary to reduce the number of required lines in this list from its end. Provided that no assumption with $C = 0$ is retained, theories gain already in confidence if the number of tentative assumptions they invoke is reduced.

## 3. Checking Standard Cosmology

In the BB paradigm, which in the late 1990s resulted in the ΛCDM "concordance model", the universe is finite in age and has emanated under conditions to which physics, as we know it, does not apply. The initial event and the primordial state belong to the fictitious domain. Our confidence in any claims that crucially depend on such an event and state cannot be any larger than zero. This does not bring the confidence in the whole paradigm down to zero, since the event is not introduced as an initial postulate but emerges as a conclusion. Concordance cosmology may still describe reality in approximation if not projected too far into the past.

It is well known that GR allows for an expanding universe and for a contracting one but not for a stationary one, unless a cosmological constant (Λ) is introduced as a means of preventing the universe from collapsing, as in Einstein's (1917a) own model of an eternal universe. Einstein had introduced Λ reluctantly, since it did not reflect anything known from physics.

Prior to the advent of BB cosmology, most natural philosophers considered the universe as eternal, but since antiquity there had been a split opinion concerning its spatial extension. According to one, the universe is spatially confined. This was still presumed by Copernicus. The competing conception of an infinite universe that perpetually regenerates itself and that contains infinitely many similar “worlds”, is also ancient. It was argued for by Epicurus, as communicated by Lucretius in *De rerum natura*, and after Copernicus by Giordano Bruno.

[2] The "PCP" needs the attribute "perfect" only because the term "cosmological principle" is in use for the imperfect cosmological principle that is respected also in GR-based cosmologies.

The first physical model of an expanding universe was presented by Lemaître (1927), who already knew that the redshift $z = (\lambda_{ob} - \lambda_{em})/\lambda_{em}$ in the light from galaxies tends to increase with their luminosity distance. According to the most straightforward interpretation of this phenomenon as a Doppler shift, the galaxies are rushing away from each other. This interpretation was adopted by Lemaître (1927), but the model was not yet a BB model. It assumed eternal expansion from an initial state, at $t = -\infty$, such as described by Einstein’s (1917a) model. In BB cosmology, $\Lambda$ was skipped, but it was reintroduced in the $\Lambda$CDM model in order to make it compatible with the magnitude–redshift relation of distant supernovae. A non-zero $\Lambda$ had already been considered earlier in order to make the age of the universe indicated by the “Hubble constant” $H$ compatible with the estimated ages of the oldest star clusters.

The interpretation of the cosmic redshift as due to an expansion of the universe is compatible with the observed redshifts, but it predicts the angular sizes of distant objects (galaxies etc.) to be larger than in a non-expanding universe.

In “tired light” models, the universe does not expand. Instead, it is assumed that light loses energy due to interaction with ingredients of the intergalactic medium or for a reason with similar effects. In the tired light model that is most often considered, since Tolman (1930), this loss causes a redshift in the light but no time dilation in its modulation. The tiredness is, thus, manifested in the carrier but not in the information conveyed by its modulation. This type of model stands falsified, since time dilation in acceptable agreement with the redshift has been observed in the light curves of distant supernovae (Leibundgut et al., 1996, Goldhaber et al., 1997, Riess, 1997, Perlmutter et al., 1998, Filippenko & Riess, 1998, Goldhaber et al., 2001, Foley et al., 2005) and in their spectroscopic aging rates (Blondin et al., 2008). However, it has never been shown that “entirely tired light”, with a time dilation consistent with its redshift would be impossible. While the label “tired light” refers to a redshift mechanism that is compatible with the Epicurean tradition, it implies otherwise no particular cosmology.

Bondi and Gold (1948) assumed their PCP to hold. Since they also considered the cosmic redshift as indicative of an expanding universe, they were led to the Steady State theory, in which creation is an on-going process by which the density in an expanding space is kept constant. This sets the Steady State theory apart from Epicurean cosmology, in which the PCP is also implied, while creation out of nothing is disallowed. Unlike the BB paradigm, which does not adhere to the PCP, and which allows models of ‘our universe’ (among other ‘universes’) to be adapted to new observations that falsify previous versions, the Steady State theory made more definite predictions. It lost adherence after the discovery of the cosmic microwave background radiation (CMB), for which it provided no convincing explanation. It can be questioned whether it ever deserved confidence, since the perpetual creation it postulates has remained as fictitious as creation in BB cosmology.

The BB paradigm also fails to provide explanations for several kinds of observational facts. In order to retain it when faced with unexpected observations, it was necessary, in the process of time, to introduce more and more free variables and fudge factors. Some of these arise directly as rational conclusions that can be drawn if the paradigm is accepted a priori. The most important were, in temporal order, 1) dark matter, 2) cosmic inflation, 3) dark energy, and 4) a particular size evolution of galaxies.

***Dark matter*** was suggested by the observed cohesion of galaxy clusters (Zwicky, 1933, 1937) and by rotation curves of individual galaxies (Rubin, Ford & Thonard, 1980). These would require much more than the visible matter to be present in order to be compatible with Classical Mechanics (CM) and GR. Initially, the hypothesis that unseen matter in form of gas, dust and substellar objects is responsible for the discrepancy was reasonable (C > 0). This matter would need to be present in haloes around galaxies and additional amounts in galaxy clusters. Since the discovery of the discrepancy, the presence of large amounts of gas in galaxies has in fact been verified, but it does not have the required mass and distribution. Neutrinos may also be considered, but the number that would be required by far exceeds the number that can be expected to have been created in a BB universe. Dark matter in form of hypothetical weakly interacting massive particles (WIMPs) is more problematic. Since attempts to verify the existence of WIMPs experimentally have so far failed, it is not justified to attach a non-zero confidence to them. They remain of type 2c in Table 1. As long as the required amount of dark matter is neither predicted on independent grounds nor empirically confirmed to be present, its supposed presence remains an excuse with $C = 0$. This means in fact that, *at the present state of our knowledge*, GR and CM stand falsified already at the scale of galaxies. Therefore, we cannot be confident in models of the whole universe based on these theories. CM actually stood falsified already when faced with the anomalous perihelion advance of Mercury and the search for the supposedly responsible planet Vulcan had failed. The problem with this single case was solved by GR. The present, more widespread and more substantial one is still awaiting its solution – which will be attempted in Section 5.3.

The galaxy rotation curves suggest that the ratio of inertial to gravitational forces is reduced for low accelerations, with a transition value of $a_0 \approx 1.1\ 10^{-10}$ m/s$^2$. This is the essence of Milgrom’s (1983) Modified Newtonian Dynamics (MoND), which allows accounting for the rotation curves of all kinds of galaxies in terms of a single function (Famaey & McGaugh, 2012). MoND also provides an explanation for the Tully-Fisher relation, which describes the otherwise unexplained close relation between luminosity and rotation velocity of galaxies. While MoND describes regularities that remain unpredicted by the dark matter hypothesis, it represents a phenomenological approach comparable to Kepler’s approach to planetary motion. It has been shown to be successful for a wide range of different galaxies. This includes galaxies with very low mass, in which the discrepancy with CM and GR is substantially larger than in the galaxies considered when MoND was originally proposed (Milgrom & Sanders, 2007, Swaters, Sanders & McGaugh, 2010). The fact that MoND describes the rotation curves of galaxies successfully in terms of a function that is at variance with CM and GR suggests that something is wrong with these theories, although MoND still requires substantial amounts of dark matter to be present in galaxy clusters (Sanders, 2003).

Among deductive approaches to MoND, two alternatives can be distinguished: 1) modified gravitation (increased where $g < a_0$) and 2) modified inertia (decreased where $g < a_0$). The theories proposed so far (Milgrom, 2002, Bekenstein, 2004) are of the first mentioned type. They involve, in addition to the Newtonian gravitational force, which varies $\propto r^{-2}$, an otherwise unknown force that varies $\propto r^{-1}$. In a different approach (Moffat, 2005), a Newtonian force combines with a Yukawa type of force instead. So far, no deductive approach to the dynamics of galaxies provides a deep understanding. Keeping GR and introducing a new, additional force whose existence has not been verified does not bring about any higher confidence than $C = 0$. Our confidence in MoND as a phenomenological model is not so low, but it involves uncertainty about the value of $a_0$ and the interpolating function between the regimes $a << a_0$ and $a >> a_0$, which counts as two free parameters.

***Cosmic inflation*** (Guth, 1981) is a purely theoretical *ad hoc* construct. It serves the explicit purpose of reconciling the fact that the universe appears flat, clumpy and yet homogeneous on the largest scale with the BB paradigm, in which such a universe would be an extremely unlikely outcome. It increases the likeliness of such an outcome by assuming physics itself to have been expediently different when the universe had not yet reached an age of $10^{-32}$ s. The whole approach and even its logical conclusiveness have long been under debate even amongst those who proposed it (Guth, Kaiser & Nomura, 2014, Ijjas, Steinhadt & Loeb, 2014). It has not been shown that cosmic inflation is anything else than a fictitious process, and these deserve no more than zero confidence. Amongst astrophysicists, cosmic inflation is not accepted by all, but this leaves the problem it is meant to solve unsolved, which does not lend any markedly higher confidence to the paradigm.

***Dark energy*** is an unpredicted fictitious form of energy with anti-gravitational properties. It is an embodiment of the cosmological constant Λ, which Einstein (1917a) introduced as a fudge factor (in form of an integration constant) when he still believed that the universe ought to be static. This Λ was reintroduced in order to make the observed magnitude–redshift relation of distant type Ia supernovae compatible with the BB paradigm (Riess, 1998, Peebles & Ratra, 2003). In the alternative *quintessence* cosmology, Λ is treated as a parameter that is allowed to vary over time (Caldwell, Dave & Steinhadt, 1998).

Assuming the existence of non-baryonic dark matter and dark energy has sometimes (e.g. Lahav & Massimi, 2014) been compared to Pauli's hesitant prediction of the neutrino, whose existence was verified only 25 years later (Mößbauer, 1998). These cases had in common that the existence of an entity that had not been known previously was suggested by abductive reasoning. However, the foundations on which these suggestions rested were epistemologically very different. The nuclear mechanism known as β-decay appeared to violate a first principle: conservation of energy. Given that this is a principle of the kind in which we can be confident even when faced with evidence that appears to contradict it, the existence of a new particle, which was later named the neutrino, was the simplest conclusion that could be drawn. This was not a fictitious assumption but a well-founded prediction. In contrast, the magnitude–redshift relation of a type of supernovae appeared to violate just the BB paradigm, in which it was not justified to be confident, and which rests on a theory (GR) that in fact stood falsified already in view of the dynamics of galaxies. The BB paradigm stands falsified also in view of this magnitude–redshift relation. Λ (dark energy) remains a fictitious excuse that lends no confidence to any reasoning about reality that is based on it.

In models in which the PCP holds, the factor by which waves are stretched per unit of distance $D$ is necessarily constant and everywhere the same. If no other mechanism contributes to the redshift $z$, we have

$$1 + z = \exp(\frac{H}{c} D) , \qquad (1)$$

where the Hubble parameter $H$ is a true constant (units $s^{-1}$ or km $s^{-1}$ $Mpc^{-1}$). In BB models, the relation is more complicated.

Inverting (1), $D$ can be calculated as

$$D(z) = \frac{c}{H} \ln(1 + z) . \qquad (2)$$

Since these expressions differ from those in the usual BB models, they predict a different relation between redshift and other observables, such as the apparent magnitude of type Ia supernovae.

In a static and flat geometry, the intensity (W m$^{-2}$) of light received from a source, flux *F* ("apparent luminosity"), varies as $F \propto D^{-2}$. *F* is proportional to absolute luminosity *L*, defined as the total power radiated by the object. If both the energy of each photon and the number of photons arriving per time unit are reduced by factors of $(1+z)^{-1}$ and if no additional factors are involved, this gives us for an object that radiates isotropically

$$F = \frac{L}{4\pi D^2 (1+z)^2} \, . \tag{3}$$

While (3) has been claimed to be valid in tired light models as well (Traunmüller, 2014), the number of photons arriving per time unit was not reduced in the casual analysis (Tolman, 1930) on which the "Tolman test" is based. In this case we get a factor of (1+*z*) in the denominator of (3) instead of $(1+z)^2$.

Recently, an analysis of redshift and magnitude data from 892 type Ia supernovae, which are the best "standard candles" we have, has shown that the two *D*'s that can be calculated on the basis of redshift (1) and flux (3) are proportional to each other (Traunmüller, 2014), so that astronomical magnitude *m* satisfies the relation

$$m = 5 \log[(1+z) \ln(1+z)] + const. \tag{4}$$

This had already been observed previously (Ostermann, 2007, Vigoureux, Vigoureux & Vigoureux, 2008)[3] in a smaller set of data that was available then. In Figure 1, $m$ - 5 log[(1+z) ln(1+*z*)], i.e., the constant in (4), is plotted against ln(1+*z*) for the individual data from Traunmüller (2014). Some variation is expected because the sources are not all of the same absolute magnitude, because of uncertainties in the measurement of *m*, and, at the lowest distances, because the proper motion of the sources can affect *z* noticeably. The observed statistical distribution is skewed, its dispersion varies to some extent, but its central tendency shows no significant variation with distance D(z).

The conclusion that the redshift factor (1+*z*) increases exponentially with distance (1) was also arrived at in an investigation (Marosi, 2014) in which the same tendency was shown to be present also in data from gamma ray bursts. The redshifts of these exceed those of the observed supernovae substantially, but it is questionable to what extent they are due to the distance of the sources. They had been taken into consideration for testing models that involve a Λ parameter (Wei, 2010).

It is clear a priori that a good fit can be obtained in Big Bang cosmology if Λ is allowed to vary as a function of time. Even if constant, Λ just describes the error of the Λ-free model, and such fudge factors lack explanatory power. The possibility of its use does not threaten the empirical validity of the simple relation (4), which follows form (1), (3) and the definition of *m*. In a ΛCDM model, the corresponding relation, $m(z;\, \Omega_M, \Omega_\Lambda)$, is more complicated and less elegant since it requires numerical integration. If neither $\Omega_M$ nor $\Omega_\Lambda$ was fictitious, an alternative that conforms to (4) directly would still be preferred because of the parsimony principle.

---

[3] I was not aware of Ostermann (2007), Vigoureux, Vigoureux & Vigoureux (2008), Marosi (2014), and Wei (2010) when submitting my paper (Traunmüller, 2014).

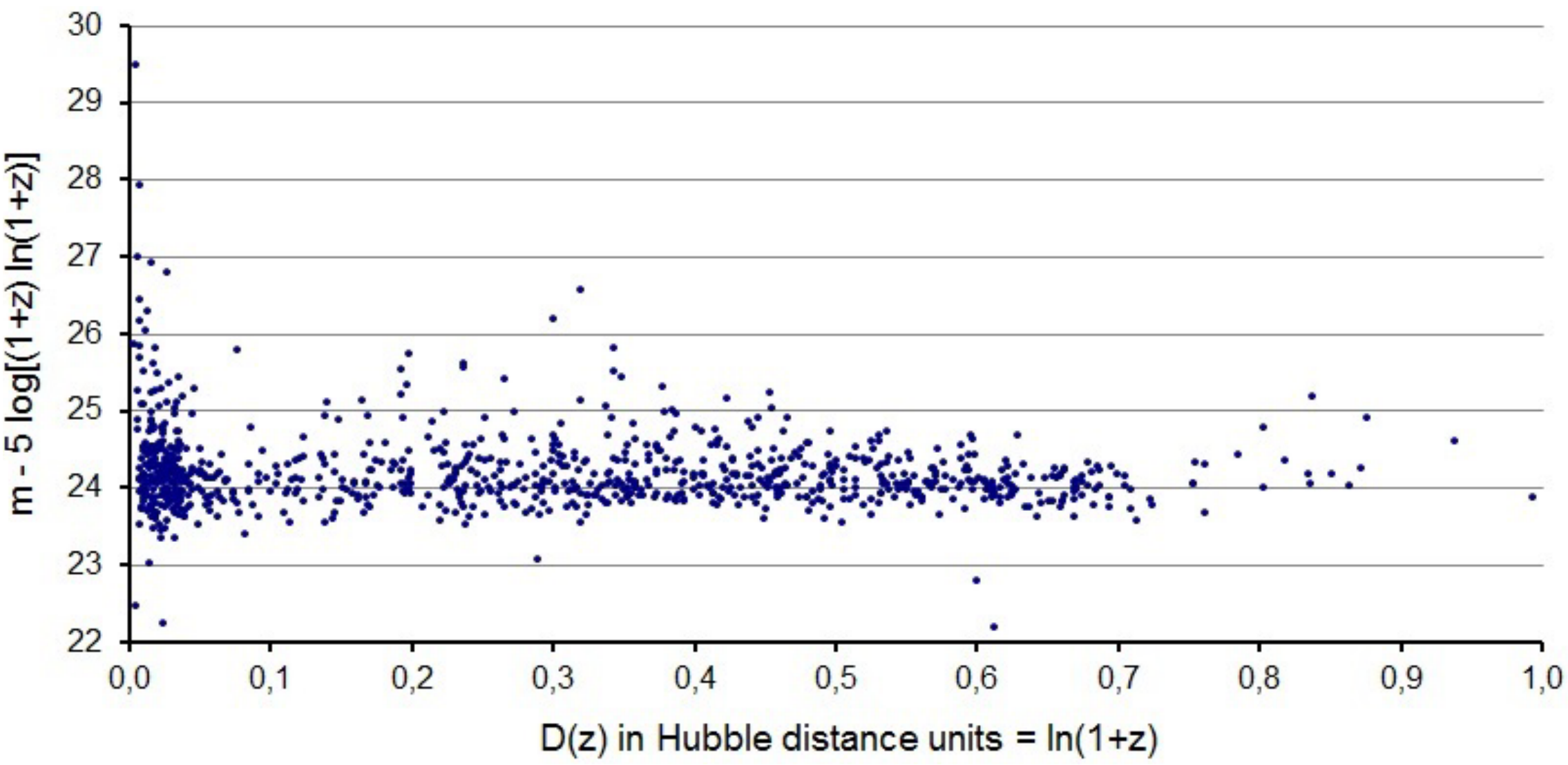

**Figure 1.** The constant in (4) suggested by each individual one of the 892 supernovae of type Ia considered in Traunmüller (2014). It is here plotted against the *z*-based distance according to (2). Its overall median is 24.13 magnitude units (mean 24.23, standard deviation 0.524).

***Size evolution of galaxies*****:** If the universe expands in proportion to a scale factor $a$ so that $a(t) = (1+z)^{-1}$, while gravitationally bound objects, such as galaxies, do not expand, the angular size $\delta$ of these will be, in small angle approximation, $\delta \approx (1+z)d/D$, where $d$ can be the major axis diameter of a galaxy and $D$ its comoving distance. This angle is enlarged by the redshift factor over that in a flat and static universe, where $\delta \approx d/D$. While tired light models predict $\delta \propto \ln(1+z)^{-1}$, all models in which the universe expands but not the galaxies, predict the relation to flatten substantially with increasing $z$ and $\delta$ to slowly increase again at large values of $z$. With exponential expansion, the prediction is $\delta \propto (1+z)\ln(1+z)^{-1}$ with a minimum for $\delta$ at $(1+z) = e$. In the past, when angular sizes were still considered to make a crucial test of the paradigm possible (Sandage, 1987, pp. 23–25), several investigations returned instead an approximate empirical relation of $\delta \propto z^{-1}$ (Sandage, 1972, Djorgovksi & Spinard, 1981). Meanwhile, measurements of the angular sizes of galaxies have progressed in scope and reliability without leading to a substantially different result (López-Corredoira, 2010). Allowing a reasonable margin for uncertainties, the observations are, instead, immediately compatible with what would be expected in a universe in which the PCP holds. If the BB paradigm is taken for granted nevertheless, this suggests that galaxies grow in size as $d \propto a(t)$ (Bowens et al., 2004, van der Wel et al., 2008, Holwerda et al., 2014) or slightly more, as $d \propto (1+z)^{-1.2}$ (Mosleh et al., 2012). It is justified to attach some confidence ($0 < C < 1$) to this suggestion, since galaxies are expected to evolve in *some* way within the BB paradigm. This evolution was formerly thought to affect mainly the luminosity rather than the size of galaxies. There is now a hierarchical theory of galaxy formation, with many free parameters (Disney et al., 2008), according to which galaxies grow by mergers of smaller pieces, dominated by dark matter. This allows modeling the empirical data, but it is an addition to the ΛCDM model – not a prediction made by it. A very extensive investigation of 4993 Lyman break galaxies ($4 < z < 10$) reported, within standard cosmology, a growth by $(1+z)^{-1.26\pm0.17}$ for the mode, $(1+z)^{-1.10\pm0.06}$ for the median and $(1+z)^{-0.95\pm0.07}$ for the mean (Shibuya, Ouchi & Harikane, 2015), which happens to be close to $a(t) = (1+z)^{-1}$.

More recently, the dark sector has been enriched by ***dark flow***. This is an observed large-scale bulk flow of galaxy clusters that appears to be in conflict with concordance cosmology. It has been tentatively ascribed to influences from pre-inflationary inhomogeneities (Kashlinsky et al., 2008, Watkins, Feldman & Hudson, 2009). Even a ***dark force***, a fifth force that affects only the fictitious kind of dark matter, has been contemplated (Farrar & Rosen, 2007). This leads deep into fairy tale physics.

ΛCDM concordance cosmology stands out as exceedingly speculative. Scientists who are not bound to the path that has led to it can easily see 1) that CM and GR stand falsified at the scale of galaxies and subsequently are not reliable at any larger scale either, and 2) that the cosmic redshift probably has been misunderstood and so given rise to fudge factors in addition to dark matter. Although even the adherents of concordance cosmology do not usually claim that they understand the universe, it is clear that by pursuing an approach that deserves no confidence, they have driven the discipline into a veritable Dark Age. When a theory persists in standing falsified, it is likely that a wrong choice has been made at a branching of the path that has led to it. In such cases, one should preferably search for the right path, instead of proceeding on the once chosen path and dreaming up an imaginary environment (WIMPs, dark energy, etc.) in which this would be the right path. Let us now find out where this wrong choice has been made.

## 4. Inertia and Gravitation

CM rests on the Galilean principle of inertia, according to which physical objects remain in their state of motion as long as no external force impinges on them. Newton took this as his first law of motion. In CM, rotation and motion in general are considered in an ‘absolute space’ and in ‘absolute time’. Most other scholars, such as Descartes, Huygens, Leibniz, Berkeley and later, Mach (1889), were of the opinion that motion can only be specified with respect to actual objects in space. Newton’s interpretation of inertia as an effect of space as such appears to have been the wrong choice that has led us into the dark. Since for a long time it met no empirical counter-evidence, it established itself as the accepted standard, but the empirical fact that the inertia of bodies is proportional to their gravitational attraction remained unexplained and intriguing.

On his path to GR, it was Einstein’s explicit objective to devise a theory in accord with Mach’s view (Hoefer, 1994). However, he actually continued on Newton’s path and consolidated it when he opened for conceiving of motion in a gravitation field as inertial motion in a curved space-time. In GR, gravitational and inertial mass are *by axiom* taken to be the same, and space has an even greater role than in CM, although it is now no longer absolute. In CM, space acts on matter, and this goes without a reaction. In GR, there is a reaction: matter curves space, but inertia remains as exceptional as in CM. Its axiomatic linkage to gravitation makes it impossible to explain the relation between the two. If a theory of gravitation and inertia is to be well-founded, it must offer such an explanation, which neither CM nor GR does.

An explanation would be at hand if the inertia of bodies would be (entirely) due to a gravitational effect analogous to electrodynamic induction in response to the relative acceleration of charges. This would be in accordance with Einstein’s (1912) conception of Mach’s principle. In this case, inertial forces would decrease when the universe expands and increase in the opposite case. They would also be increased in the vicinity of a gravitating body, where distant bodies appear blue-shifted and, in effect, closer by.

An attempt to develop a theory in which the inertial force is induced by the relative acceleration of the masses of the universe is due to Sciama (1953). He described a vector theory as a step towards a tensor theory compatible with GR. While he investigated some of the cosmological consequences of his theory, he did not devise a cosmology from scratch on the altered premise. He assumed the universe to be expanding, in accordance with the prior interpretation of the cosmic redshift. In attempting to account for inertia on the basis of the density of the universe inferred from astronomical observations, he obtained a missing mass problem of the same magnitude as in GR-based cosmological models, but he expected large amounts of uncondensed and yet unobserved matter to be present between galaxies.

In order to be tenable and useful at the scale of galaxies and above, a theory of gravitation and inertia must also explain the dynamics of these. As a phenomenological model, MoND does this only at the most superficial level. It allows predicting the rotation curves of different galaxies and so can be said to explain the relations between them, but it does it in the absence of an understanding of the underlying physics. The physically founded theories proposed so far (Milgrom, 2002, Bekenstein, 2004, Moffat, 2005) attempt to improve this, but since they introduce an ad hoc force ($C = 0$) of some kind, they actually lack epistemic value. They are also instances of proceeding on the established path and “dreaming up” something that would make it right. The alternative would be to reconsider the path that was suggested, but not elaborated, by Mach (1889) and not fully appreciated by Sciama (1953), and to see where it leads. This will be undertaken in Section 5.3.

## 5. In Search of a Tenable Cosmology

### *5.1. Primary Considerations*

If cosmology is to be an empirical science, it is a minimum requirement that its tenets impart more than zero confidence. The theory must not assume any unpredicted fictitious entities or processes to be in effect. Ideally, it should be based on definitions and first principles alone. In addition to these, only the most generalizing assumptions shall be accepted. To these belongs the PCP:

*The universe is homogeneous and isotropic in time as well as in space.*

Such a universe is persistent instead of transient. Its statistical properties do not change as a function of time, space, and direction. The PCP should, however, only be assumed to hold within volumes that are sufficiently large – that of a Hubble sphere or larger. Up to distances of at least 200 Mpc, the observed distribution of matter is actually far from homogeneous – it is rather fractal (Baryshev et al., 1998).

Further, gravitation is to be treated in such a way that it allows inertia to emerge as an effect of the gravitation of cosmic masses (Mach’s principle). This, as well as the PCP, is at variance with GR, which is actually promising, considering that GR is incompatible with quantum mechanics and fails to account for the rotation curves of galaxies.

The PCP puts a narrow constraint on the redshift–distance relation. Abstracting influences of nearby masses away, the function must be self-similar and the same everywhere in space-time: in a flat geometry, this can only be a constant exponential function $(1+z) \propto \exp(D)$, so that (1) and (2), which, in addition, only contain the constants $H$ and $c$, must hold. Further, if extinction of light is negligible or compensated for, (3) and (4) must also hold. This is in adequate agreement with the empirical flux–redshift (and magnitude–redshift) relation of supernovae SN1a (Traunmüller, 2014), which corroborates the tenability of the PCP.

**Table 2.** Some explananda of cosmological theories.

| **Phenomenon** | **Expression** | |
|---|---|---|
| Time dilation factor | 1+z | |
| Flux–redshift relation and magnitude–redshift relation | $F \propto [(1+z)\ln(1+z)]^{-2}$,<br>$m = 5\log[(1+z)\ln(1+z)] + const$ | (5)<br>(4) |
| Angular diameter of galaxies | $\delta \propto (1+z)^{\alpha} \ln(1+z)^{-1}$, with $\alpha \approx 0$ | (6) |
| Cut-off acceleration $a_0$ of galaxies (in MoND) | $0.13 < a_0 c^{-1} H^{-1} < 0.22$ | (7) |

In order to be tenable, a cosmological model must account for the four explananda in Table 2. Unfudged BB models account for time dilation but fail for the other three. Cosmologies with exponential expansion, such as the "Scale Expanding Cosmos theory" (SEC) (Masreliez, 2004) and the model by de Sitter (1917)[4] in addition account for the flux–redshift relation, but the latter model has the blatantly fictitious property of containing no mass. The observed angular sizes of galaxies are at variance with the distance-duality relation. This relation (Etherington, 1933) is said to hold if photons travel along null geodesics in a Riemannian geometry, and their number is conserved, but it becomes practically inapplicable if galaxies evolve in luminosity or size with $a(t)$. If they do not, luminosity distance $D_L$ is related to angular distance $D_a$ as

$$D_L = D_a(1+z)^2. \qquad (8)$$

This would require $\alpha = 1$ in (6). A violation of the distance-duality relation has been observed between $D_L(z)$ of supernovae Ia and $D_a(z)$ of radio galaxies, compact radio sources and X-ray clusters (Bassett & Kunz, 2004). Analyses of data from galaxy clusters have more recently been reported to be compatible with the relation (8), the elliptical model fitting better than the spherical (Holanda, Lima & Ribeiro, 2011), or vice versa (Yang et al., 2013). However, these papers are concerned with the relation between $D_a(1+z)^2$ and $D_L$ in (8) and do not report an estimate of the crucial exponent α in the first term of (6). The result of the investigation by Shibuya et al. (2015) is similar to those of previous investigations, which all suggest this exponent to be closer to 0 than to 1, but the difference between 0.05 for the mean, –0.10 for the median and –0.26 for the mode is in need of an explanation. A failure to consider the intricate effect of gravitational self-lensing of galaxies on their measured angular sizes may be involved here.

The universality principle suggests that Planck's radiation law, i.e., the Stefan-Boltzmann law and Wien's displacement law, should be valid for sources at any distance. This allows for the surface brightness *SB* of a redshifted blackbody a maximum of $SB \propto (1+z)^{-4}$. In the absence of extinction, lensing and redshift, *SB* does not change with distance. If there is only a redshift, so that flux accounts for a reduction by $(1+z)^{-2}$, the solid angle the object subtends must increase by a factor of $(1+z)^2$. This condition is satisfied if the distance-duality relation holds, but it is not always evident that it does. In the case of galaxies, there can hardly ever arise a conflict, since their *SB* is much lower than that of a black body. In the case of stars in distant galaxies, the problem remains an academic one, since these are bound to remain point-like sources for which *SB* cannot be measured.

[4] This model is said to evolve as $a(t) \propto \exp(Ht)$, but in de Sitter's original conception, there was no real expansion: "*the frequency of light-vibrations diminishes with increasing distance from the origin of co-ordinates. The lines in the spectra of very distant stars or nebulae must therefore be systematically displaced towards the red, giving rise to a spurious positive radial velocity*".

The cut-off acceleration $a_0$ is the most challenging one of the explananda in Table 2. Its empirical boundaries correspond to $cH$ = 6.1±1.5 $a_0$. This should preferably not be a free parameter, which it is in MoND, but emerge from well-founded cosmological considerations.

### *5.2. The Delimitation Problem*

Within GR based cosmology, galaxies are thought of as essentially remaining at rest in an expanding space that brings light waves to expand with it. Friedmann-Robertson-Walker (FRW) models and other GR-based alternatives describe the relation between space-time affected by gravitation and expansion, and space-time in absence of these influences, e.g. in the equation for the line element. Standards of comparison, such as sources of radiation and the meter, which is defined in terms of light waves, are, however, tacitly exempted from these influences. The proper space-time defined by them is not identified with the space-time influenced by the potential and expansion at their place, but with one that is free of such influences. This is logically in error but appears empirically right, because it immediately predicts an observable cosmic redshift. A logically error-free treatment requires a criterion for delimiting what expands from what does not. FRW models as such do not offer such a criterion. There are proposals, such as the so called Swiss cheese models (Harwit, 2006), which allow to model regions in which different metrics apply, but standards of comparison are still tacitly exempted even in these. A solution of the delimitation problem does not require a more refined metric but a rule for when the metric that is valid in the absence of gravitation and expansion is to be used instead – if this should be logically defensible at all.

The delimitation problem is avoided (it does not arise) in the common pseudo-Newtonian approach, in which "expansion" is an attribute of structures and distances rather than of "space" in a substantivalist sense. In this approach, the expansion takes effect only to the extent to which it is not prevented by forces. In this matter, gravitation is not treated differently from other forces. Thus it is, essentially, assumed that

(Criterion 1) *free waves and incoherent objects expand – coherent objects do not.*

In such an approach, it has been calculated (Giulini, 2014) that expansion sets in at a radius of 10 Mpc for a structure like the Virgo cluster, which comprises more than 1000 galaxies within a radius of less than 3 Mpc. If the coherence extends this far, crit. 1 leaves only the larger voids between galaxy clusters to expand.[5] However, one can doubt whether even these could expand. The universe looks like a three-dimensional web of galaxy clusters, which are connected to their neighbors via filaments whose matter density appears to suffice for coherence along their axes. Persistent filaments would prevent even the voids from expanding. It is, at least, clear that the regions that now cohere would have overlapped when the BB universe was younger, which would have prevented its expansion alltogether. Therefore, this approach fails to offer a workable delimitation between what expands in a BB universe and what does not. As discussed under "Size evolution of galaxies" in Section 3 and by López-Corredoira (2010), the observed angular sizes of galaxies, (6), are also hard to reconcile with an overall expansion. However, crit. 1 remains compatible with a non-expanding universe, in which any signals that propagate at *c* expand, while the infinite cosmic web is, essentially, static.

---

[5] The pseudo-Newtonian approach has a problem with recession velocities > *c*, which require switching to a substantivalist space again, but the velocities involved here remain << *c*.

It is also possible to consider the expanding entities of crit. 1 as static while coherent objects and structures shrink and processes speed up (Traunmüller, 2014, Arora, 2015). In short,

(Criterion 1b) *coherent objects contract – free waves do not.*

With crit. 1, the expansion is exponential; with crit. 1b, it is exponential in reciprocal proportion. Criterion 1b describes the same situation as crit. 1 in a frame of reference that is co-expanding with the waves. Any observable effects are the same. If the cosmic web is coherent, there is no observable expansion or contraction of objects of any size. The entities that contract in the contraction model (1b) include any real standards of comparison. The material universe remains, therefore, metrically static and so compatible with the PCP[6]. The metric space being the same, this brings us back to the expansion model with crit. 1.

Within the frame of GR, on which the FRW models are based, crit. 1 appears incongruous since GR does not draw a distinction between coherent and incoherent objects. Instead, it draws a distinction between non-gravitational forces and gravitation, linking the latter directly with space. This link is broken in crit. 1 by exempting gravitationally bound objects from the expansion. GR rather suggests delimitation between a space in which radiation propagates, and which is also the space of gravitation, and the space of non-gravitational forces, which can be equated with that of CM. A corresponding assumption would be that

(Criterion 2) *anything under free gravitation expands – objects under control of other forces do not.*

This alternative predicts the universe, the cosmic web, galaxies and planetary systems all to expand, which is incompatible with the PCP but not necessarily with the observations. The angular size discrepancy may disappear if galaxies participate in the expansion. Further, if planetary systems expand, this would be reflected in an increase of the Astronomical Unit (AU). With AU = 149.6 $10^9$ m, and $H$ = 60 km $s^{-1}$ $Mpc^{-1}$, there would be a secular increase by 17.8 m for expansion by (1+z). A secular increase of the AU by 15±4 m has actually been reported to be present in empirical data (Krasinsky & Brumberg, 2004). Essentially the same explanation might also account for the increasing eccentricity of the lunar orbit (Iorio, 2011). However, Pitjeva & Pitjev (2012) reported a non-significant increase of the AU by only 1.2±3.2 m per century (at the 3 σ level).

As for the rotation curves of galaxies, crit. 2 amplifies the discrepancy with the astronomical observations. Aside from this trouble, it does very well if it is true that planetary systems and galaxies participate in the general expansion, or perhaps just in the expansion supposedly caused by dark energy (roughly 50%) (Křížek & Somer, 2013). The efficacy of crit. 2 can be falsified by demonstrating that the AU does not increase correspondingly. The reports of its increase (Krasinsky & Brumberg, 2004, Iorio, 2011) or absence of significant increase (Pitjeva & Pitjev, 2012) were based on results that are highly sensitive to small errors of various kinds.

---

[6] Traunmüller (2014) described such a model (not fully understood) as making different predictions from expansion models. However, if the cosmic web does not expand, correct predictions based on (1) and (1b) will agree.

An error that might feign or hide a change in the AU appears to be the cause of the "Pioneer anomaly" (Anderson et al., 2002, Turyshev & Toth, 2010). This is an unexplained acceleration of about 8.7 $10^{-10}$ m/s$^2$ directed towards the Sun, observed in the trajectories of space probes. While the anomaly was explained away as a thermal effect (Turyshev et al., 2012), it rather reflects an erroneous modification in the acquisition or processing of data that was introduced in 1990. This is evident from an exercise in which publicly accessible data were analyzed in order to verify the anomaly (Unzicker & Schmidle, 2007). The graphs in that paper show that there was no anomalous acceleration before a certain date, when it suddenly appeared and remained in the data from both Pioneer 10 and 11 (launched 13 months later). This goes unmentioned in the cited papers (Anderson et al., 2002, Turyshev & Toth, 2010, Turyshev et al., 2012, Unzicker & Schmidle, 2007). Since the mistake, perhaps GR related,[7] turns up in the tracking of at least two space probes and remains there over the years, it may be present also in data that have been used in investigations of the constancy of the AU (Krasinsky & Brumberg, 2004, Iorio, 2011, Pitjeva & Pitjev, 2012). This needs to be cleared up in order to judge whether crit. 2 is tenable or stands falsified if the PCP is allowed to be violated and the dynamics of galaxies to be left an open problem.

If the PCP holds and the cosmic web is coherent, crit. 1 implies that the material universe is 'static' on all scales, while waves are stretched. If the factor by which waves are stretched per unit of distance is constant and everywhere the same, the redshift factor (1+$z$) increases exponentially with distance $D$. If the number of periods between a source of radiation and the observer is conserved, which is the case in "entirely tired light" models, the expanded distance $D_{\mathrm{exp}}$ can be calculated by integration as

$$D_{\mathrm{exp}} = \exp(\frac{H}{c}D) - 1\,. \tag{9}$$

Under this condition, $D_{\mathrm{exp}}$ is simply proportional to $z$,

$$D_{\mathrm{exp}} = \frac{c}{H}z\,. \tag{10}$$

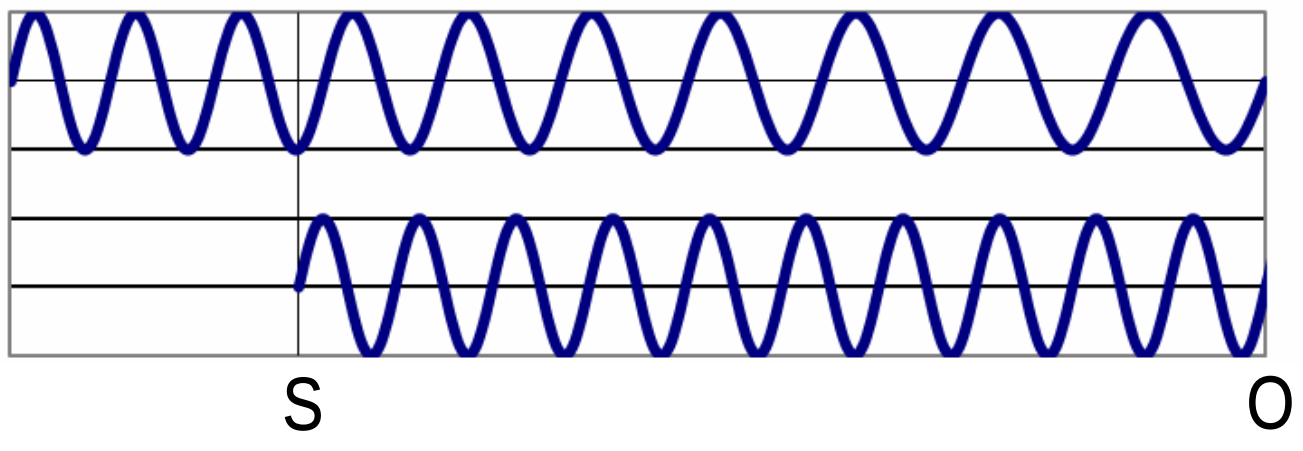

**Figure 2.** An unexpanded wave train (below) from source S to observer O at distance $D$ = 0.5 Hubble length units ($cH^{-1}$) and its expanded equivalent (above). The chosen $D$ gives a redshift $z$ = 0.649 and an expanded distance $D_{\mathrm{exp}}$ = 0.649 $cH^{-1}$ (1.297 $D$).

[7] In Turyshev & Toth (2010) one can read: "*One can demonstrate that beyond 15 AU the difference between the predictions of Newton and Einstein are negligible*". This is said without telling that it holds only for observations made from a still larger distance – not for those made by us from Earth. This evokes a suspicion that the Pioneer anomaly may have arisen from a similar inadvertence.

The expansion is illustrated in Figure 2. $D_{exp}$ is a distance that is valid for signals that propagate at *c*. Since equations (9) and (10) hold irrespective of frequency, down to zero, they also hold for the effective lengths of lines of force. It is noteworthy that the unexpanded distance 2*D* from a source to a mirror and back can theoretically be measured by counting the periods of a stable monochromatic signal that can be sent towards the mirror until the first period of the reflected signal returns. No period is lost, and the signal propagates at *c*, while the distance it appears to cover expands together with the wave. Not only waves but any field modulations that propagate or are maintained at *c* are dilated in this way, whereby all slopes and gradients become successively smaller.

Crit. 1 is not explicit about the reason for the dilatation, but one can easily see a reason for waves to expand if one considers *c* as the maximum velocity of non-escape from the universe, which is tantamount to an *explanation* of *c*. Under this premise, anything that moves at *c* will have to overcome an omnipresent non-zero gradient: it is pulled back by the gravitational attraction of the universe, which in Section 5.3 will be shown to be finite. This gradient is nearly insensitive to inhomogeneity in mass distribution such as observed in the universe and which needs to be considered if the Friedmann model is to be applied (Baryshev et al., 1998). It will be disproportionately smaller for anything that moves at $v < c$, when the distant masses no longer pull in the same direction. Such cases remain outside the frame of the present paper, which, in addition to objects that move at $v << c$, is primarily concerned with waves and signals that propagate exactly at *c*. For these, (9) and (10), illustrated in Figure 2, are valid.

The two delimitation criteria are contrasted in Table 3. For both, a contraction model has been entered in addition to the equivalent expansion model. One is free to choose one of the two ways of regarding the situation. There is no such freedom if fudge factors are introduced in the way this is done in ΛCDM cosmology, and the two ways result in equally simple descriptions only if expansions/contractions are exponential functions.

**Table 3.** Delimitation criteria of cosmological models in which expansion is exponential as in (1) and (2), flux $F \propto [(1+z)\ln(1+z)]^{-2}$ (5). The entities that contract in contraction models are those that remain unaffected in expansion models.

| **Delimitation criterion** | **Expansion models**<br>Entities that expand | **Contraction models**<br>Entities that contract |
|---|---|---|
| **1 Coherence** | *Free waves and incoherent objects*<br>(radiation, slopes and gradients propagating or maintained at *c*; universe if incoherent) | *Coherent objects*<br>(atoms, rocks, planets, stars, galaxies, galaxy clusters; cosmic web if coherent) |
| **2 Gravitation** | *Anything under free gravitation*<br>(waves, universe, cosmic web, galaxies, planetary systems) | *Objects under control of non-gravitational forces*<br>(atoms, rocks, planets, stars) |

Crit. 1 reflects the practice in BB cosmology, which, as we have seen, appears to be incompatible with an expanding material universe. It is, however, compatible with a cosmology in which the PCP holds and which, unlike CM and GR, offers an explanation for inertia, as detailed in Section 5.3. It appears also to be compatible with the empirical data unless the AU actually increases as it would according to crit. 2. This needs yet to be checked empirically.

Crit. 2 is in the gist of GR, in which gravitation is special by being linked to space. It implies an expanding universe in which also planetary systems expand, but the empirical flux–redshift relation (5) suggests the expansion to be exponential, as it is in SEC (Masreliez, 2004), in which, however, the delimitation criterion has not been made explicit either. In the de Sitter universe (1917), any measurements are imaginary since it contains no matter, but this is not much different in FRW models, which, strictly speaking, do not either contain any of the aggregations of matter, such as atoms, instruments, planets, stars and galaxies, but only an abstract fluid. We shall restrict the further discussion to models in which crit. 1 applies.

Even if a contraction model (crit. 1b) is mathematically equivalent to an expansion model (crit. 1), there is a heuristic difference between the two. Model 1 suggests the mentioned explanation of the redshift. This is hidden in model 1b, which discloses no direct explanation for the contraction of coherent objects either. On the other hand, model 1b makes it obvious that there must be time dilation in proportion to the redshift factor. This is hidden in model 1, which might, mistakenly, be thought to lack overall time dilation.

### *5.3. Gravitational Potential, Gradients and Inertia*

In GR, a static gravitational force and a force due to uniform acceleration of a body have been made equivalent, in accordance with Einstein's (1907, 1911) equivalence principle, by treating not only inertia but also gravitation as an action of space. One can, alternatively, reason like this:

*If the force that acts on a body at rest on Earth is given by a gradient in the gravitational potential field of the Earth, the force that acts similarly on an accelerated body must then be given by a gradient in a field that is present in the comoving frame of the accelerated body.*

In this comoving frame, the rest of the universe is seen as accelerating in the opposite direction and must so give rise to a force in this direction. This force must be counterbalanced in order to accelerate the body. This alternative embodiment of the equivalence principle implements Mach's principle immediately.

Under ordinary conditions, the inertial force is $\boldsymbol{F} = m\boldsymbol{a}$. The dynamics of galaxies could then possibly be explained if it could be shown that $F$ is reduced disproportionately for accelerations that are not much larger or even smaller than Milgrom's $a_0$.

Let us first consider the scalar gravitational potential $\Phi$ that is due to all the masses of the universe. It can be calculated for any point in space-time by summing up the contributions from all masses $m$ at their distance $r$ from the point,

$$\Phi = -G\sum \frac{m}{r} . \quad (11)$$

In a homogeneous non-expanding universe, this $\Phi$ comes out as $-\infty$, which leads to absurdities. However, in the suggested cosmology, the effective $r$ in (11) is not the static distance $D$ but the expanded distance $D_{\text{exp}}$ of (9) and (10). With this distance, we get

$$\Phi = -G\sum \frac{mH}{cz} . \quad (12)$$

In a universe in which matter is homogeneously and isotropically distributed in static space, the potential can be calculated by integrating the contributions from shells of thickness d$r$ at distance $r = 0$ to $\infty$:

$$\Phi = -4\pi G\rho \frac{H}{c}\int \frac{r^2}{z}\,\mathrm{d}r. \qquad (13)$$

The contributions to $\Phi$ by shells up to $r$ = 8 $cH^{-1}$ in (13) are shown in Figure 3 (continuous line). The integrated contributions (from $r$ = 0 to ∞) are 4.80823 times larger than those calculated for a sphere without expansion and $r_{\mathrm{max}} = cH^{-1}$.

It is commonly claimed that the ordinary baryonic matter accounts for no more than 4.5% of the critical density, $\rho_c = 3H^2(8\pi G)^{-1}$, of a BB universe or a Newtonian Hubble sphere. Actually, only about 0.5% of $\rho_c$ is accounted for by matter that can be said to have been "observed" in stars and galactic clouds. The remaining 4% is believed to be present in warm intergalactic plasma whose estimate "is driven by the need to balance the budget rather than more directly by the observations" (Fukugita and Peebles, 2004). Even if the potential of the universe is 4.8 times larger than that of a Hubble sphere, baryons will still only contribute relatively little to the total. The major contributions may come from neutrinos and electromagnetic radiation (photons), whose original energy is not lost as their redshift progresses, but transmuted into gravitational form. This is briefly discussed in Section 5.4.

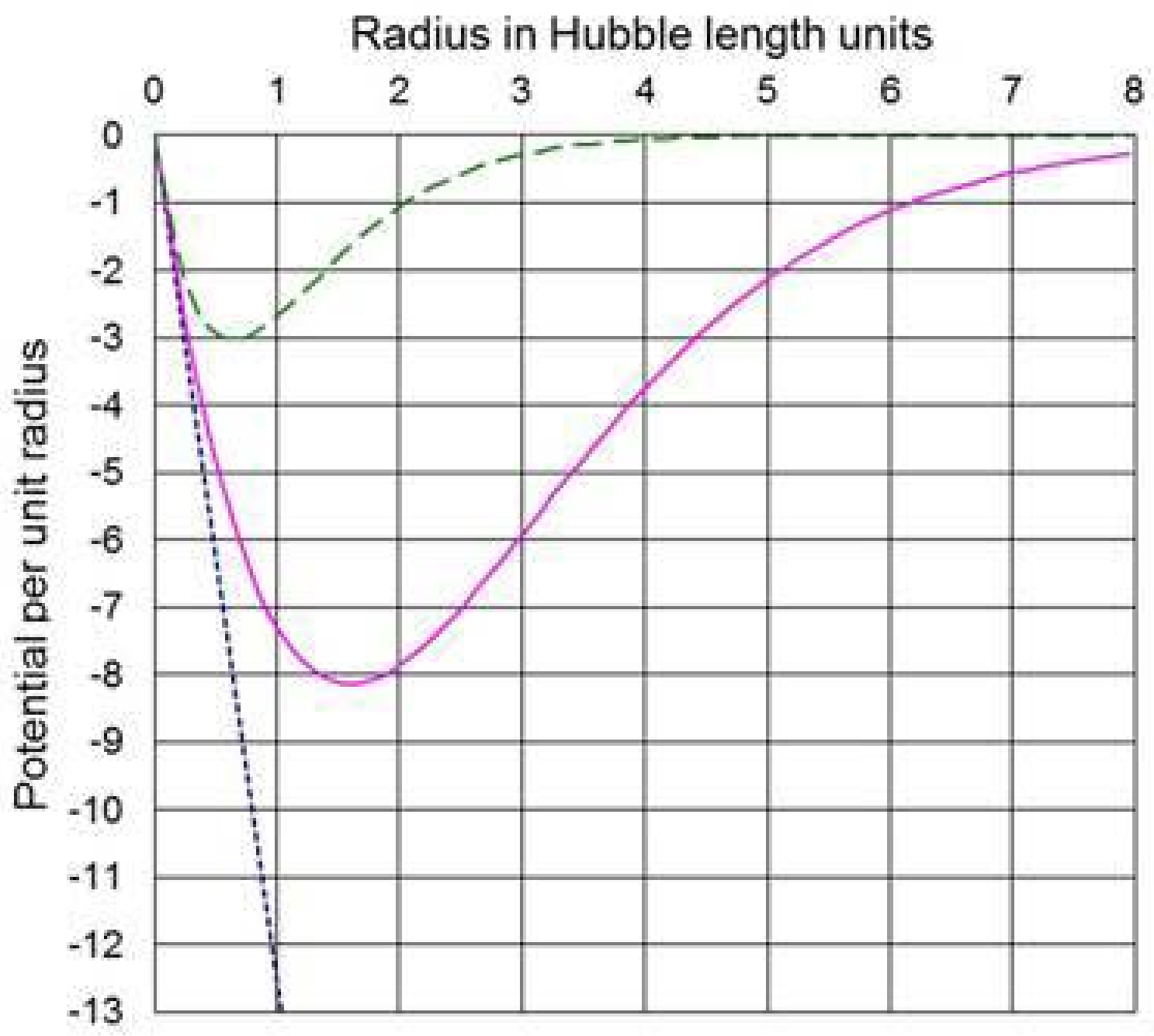

**Figure 3.** Potentials per unit radius in a homogeneous isotropic universe shown as a function of the radial distance (in Hubble length units $cH^{-1}$) from an observer. Naive contributions to $\Phi$ [as if (11) was valid] (dotted line) and those in a cosmology in which (13) is valid (continuous line). These are less negative by the factor $D/D_{\mathrm{exp}}$. The dashed line shows the contributions to the equivalent potential $\Phi_{equ}$ of (15), which are less negative by an additional factor of $(1+z)^{-1}$.

The Hubble acceleration $cH$ describes a dilatation (a stretching action) by which the slopes and gradients of gravitational signals are reduced isotropically. This is an effect that becomes preponderant only at accelerations that are still smaller than the small $cH$. The acceleration $\boldsymbol{a}_{\mathrm{red}}$ that corresponds to a gradient that is reduced by this action is

$$\boldsymbol{a}_{\text{red}} = \frac{\boldsymbol{a}}{1 + \frac{cH}{a}}. \tag{14}$$

If the inertial force that needs to be overcome in order to impart an acceleration $a$ on a body is entirely due to the acceleration of the rest of the universe in the opposite direction, we might, thus, expect this force to be reduced like $\boldsymbol{a}_{\text{red}}$ in (14). Observations tell us that there is such a reduction, but that the cut-off acceleration $a_0$ (Milgrom, 1983) is still lower than $cH$. This can be understood and explained as follows: If everything but a test body accelerates uniformly in the same direction, the acceleration of a distant mass 'seen' by the test body is not $\boldsymbol{a}_{\text{red}}$ (which is reduced already at the origin) but its dilated equivalent, $\boldsymbol{a}_{\text{red}}(1+z)^{-1}$, which is seen from a distance. This suggests that Milgrom's $a_0$ is an averaged dilated view of the Hubble acceleration $cH$. While the gradient that corresponds to $\boldsymbol{a}_{\text{red}}(1+z)^{-1}$ is due to interaction with all the distant masses, it can properly be considered to have its origin in the accelerated body. The information propagates from there and not from distant objects that were accelerated billions of years ago.

In order to calculate $a_0$, we have to find the weighted mean value of the factor $(1+z)^{-1}$ by which accelerations appear reduced from a distance, and to multiply it by $cH$. The weighting must be proportional to the contributions to $\Phi$ by each shell, see (13) and the continuous line in Figure 3. The dashed line in the figure shows the so weighted, less negative contributions to an equivalent potential

$$\Phi_{\text{equ}} = -4\pi G\rho \frac{H}{c} \int \frac{r^2}{z(1+z)} \mathrm{d}r. \tag{15}$$

In Figure 3, $\Phi_{\text{equ}}$ is represented by the area between the abscissa and the dashed line. It is found to be smaller than $\Phi$ (the area between the abscissa and the continuous line, extended to $r = \infty$) by a factor of 0.168093 ($\Phi/\Phi_{\text{equ}} = 5.94910$). The same result is obtained by calculating the mean $r$ of the distribution shown by the dashed line and finding the value of $(1+z)^{-1}$ for this $r$. We get $a_0 = 0.168093\ cH$.

If the inertial force goes towards $\boldsymbol{F} = m\boldsymbol{a}$ at $a \gg a_0$, is given by a gradient in the field seen by an accelerated body, and slopes are dilated in the way described, the equation for the inertial force becomes

$$\boldsymbol{F} = \frac{m\boldsymbol{a}}{1 + \frac{0.168093\ cH}{a}}. \tag{16}$$

This reduced inertial force appears to explain the observed galaxy rotation curves and their successful description by MoND.

Since MoND does not fix the interpolating function between the regimes $a \ll a_0$ and $a \gg a_0$, several such functions have been tried (Famaey & McGaugh, 2012, Famaey et al., 2007). Equation (16) singles out the "simple" interpolating function as the predicted one, with $1+a_0/a$ in its denominator. This function was actually reported to give a better fit to empirical data than the "standard" interpolating function, whose preference derives from the fact that it approaches the Newtonian law more closely at $a \gg a_0$ (Famaey et al., 2007). It would require $[1+(a_0/a)^2]^{1/2}$ instead of $1+a_0/a$ in (16). The two interpolating functions do not give the same optimal value for $a_0$, which lies in the range of $cH/a_0 = 6.1\pm1.5$, but the value of $cH/a_0 = 5.94910$ implied by equation (16) falls well within this range. It remains to be investigated how well equation (16) actually fits the observational data.

There is, in addition, a small effect of expansion on the static gravitation of galaxies, which is marginally reduced in their outskirts. This can be taken into account by substituting *cz/H* for *r* in the calculations, as we have done in going from (11) to (12). If we consider only galaxies and, at most, their satellites, the difference between *cz/H* and *r* remains small enough to be neglected here.

In distinction from MoND, the present approach is deductive and based on a well-founded alternative to Newtonian Mechanics. While it needs to be reflected on and elaborated more deeply, it requires no ad hoc assumption. It requires just skipping Newton's questionable assumption about the cause of inertia and considering the consequences within the frame of a field theory. If such a non-speculative approach works, it is clear that fundamental progress has been made.

While the validity of equation (16) does not depend on the composition of the energy content of the universe, the equation needs to be modified in order to capture the effect of a local gravitational time dilation. Gravitational time dilation was predicted (Einstein, 1907) already before GR had been elaborated. Nowadays, its existence can be demonstrated directly with precision clocks at different heights in the gravitational field of the Earth. It causes an isotropic blueshift of light that arrives from distant sources. If inertia is due to the acceleration of the universe in the rest frame of an accelerated body, the slopes that communicate this acceleration are also "blueshifted", i.e., they appear steeper. Thereby, the inertial force $\boldsymbol{F}$ is increased. Defining the gravitational blueshift $z_g$ in harmony with the cosmic redshift as $z_g = (\lambda_{ob} - \lambda_{em})/\lambda_{em}$, it will be in the range $0 > z_g > -1$. With this we get

$$\boldsymbol{F} = \frac{m\boldsymbol{a}}{1 + z_g}. \tag{17}$$

Equation (17) implies that "black holes", in which a source of gravity is surrounded by an event horizon at which $1 + z_g = 0$, cannot form, since the force $\boldsymbol{F}$ would go to infinity there. This is at variance with GR as commonly taught, but not with any actual observations.

The combination of (16) and (17) finally results in

$$\boldsymbol{F} = \frac{m\boldsymbol{a}}{\left(1 + z_g\right)\left(1 + \frac{0.168093\, cH}{a}\right)}. \tag{18}$$

Possibilities for checking the validity of (18) by observations within the Solar System remain to be contemplated.

### *5.4. The Cosmic Energy Cycle*

The statistical properties of a persistent universe remain constant over time. This raises a range of important questions to which hardly any attention has been paid before, since they do not arise in a transient universe. They concern the recycling of energy via a cosmic ocean of photons (CMB), gravitons and neutrinos, which interact with each other in ways that need to be studied. If energy is conserved, which is a first principle, the energy that is 'lost' due to the cosmic redshift can only be shifted to another carrier and/or place in a cosmic energy cycle (Edwards, 1998). Since this is not so in a transient universe, the topic is never touched in BB cosmology, while it is of central importance if the PCP holds.

Edwards (2012) considered the possibility that the energy that photons loose by becoming redder might be directed towards gravitation, making gravitational potential wells deeper.

Provided that the same holds for neutrinos and gravitational radiation as well, all the energy that stars radiate away would so be recycled back to stars and other heavenly bodies. Such a process is also apt to prevent a “heat death” of the universe. Edwards (2012) did not develop a theoretical foundation for his hypothesis, but he estimated the implied energy turn-over in the opposite transition of the cycle, where energy is assumed to be transferred from gravitons to photons, in different types of astronomical objects.

The scenario described in Section 5.3 is compatible with Edwards’ hypothesis. Stars loose energy and mass when they emit neutrinos and photons. When this radiation is redshifted as it propagates through the universe and has to overcome a gradient, its energy content is gradually converted into gravitational radiation. This gravitational radiation flows into a cosmic ocean that defines a floor for the redshift. It contains mainly gravitons, which interact gravitationally with neutrinos and photons. The gravitons are eventually absorbed in potential wells, which so become deeper unless the effect is balanced (or overridden) by emission of radiation, which makes the wells shallower. Together, the bodies in the universe absorb as much energy, mainly in gravitational form, as they radiate per unit of time, mainly in form of neutrinos and photons. While details remain to be worked out, we can be sure that the 2.725 K of the CMB is the blackbody temperature at which this balance is obtained. Given the PCP, this temperature is bound to remain constant.

The CMB photons account for a fraction of 5 $10^{-5}$ of $\rho_c$ (Fukugita and Peebles, 2004). They must have their origin in starlight, no other choice being offered here. Let us assume that they have been redshifted, on average, by a factor of 1600 since they left the surface of last scattering (at 4360 K). In this case, we can expect the energy present in form of gravitons to be 1599 times higher than that present in form of CMB photons. This would sum up to 8% of $\rho_c$. Further, stars emit about 1.6 times as much energy in form of neutrinos as they emit in form of photons (Fukugita and Peebles, 2004), more of the neutrino energy being due to core collapse than to nuclear burning. This would, then, already bring us up to the 20.8% of $\rho_c$ that appear to be required in a universe in which (13) is valid. Caveat: this is just a check of reasonability - a more careful investigation of the energy content of the universe exceeds the scope of this paper.

In attempts to explain the excessive cohesion of galaxy clusters, it has been suggested that massive neutrinos might be responsible (Sanders, 2003, Swaters, Sanders & McGaugh, 2010). In a BB universe with its limited age, this requires the presence of primordial or extra massive neutrinos ($\approx$ 2 eV instead of $\approx$ 0.05 eV). In a universe with an infinite past with essentially unchanging properties, there is no shortage of neutrinos produced in known processes, and it will, on average, take many Hubble times until a neutrino will be captured again. At that point, its rest mass may account for most of its remaining energy content, so that it will move at $v \ll c$ and be captured by a galaxy cluster. This might provide a non-ad hoc explanation.

## 6. Discussion

Although a fully elaborated more well-founded cosmological theory remains yet to be presented, a path along which such a theory can be arrived at has been identified in the preceding sections and shown to be easily passable and worthwhile to follow. It leads to a physical cosmology that is more in accord with the ancient world view of Epicurus (ca. 341–270 BC), Lucretius (ca. 99–55 BC) and Giordano Bruno (1548–1600) than with the one arrived at by proceeding on the path indicated by Newton and followed by Einstein. Nature has already told us clearly enough that this is the wrong path.

The non-speculative cosmology that emerges represents a Machian alternative to GR whose predictions deviate from those of GR both where gravitation is very weak and where it is very strong, as can be seen in (18). GR will remain a limiting case of such a more comprehensive theory, to which it points out the way, and Einstein (1917b) considered this to be the fairest destiny a physical theory can have.[8] However, there is a limit to such developments: the most well-founded theories can no longer be topped in this way, since they will themselves be the most comprehensive ones.

It remains to be checked whether other observables than the most basic ones, listed in Table 2, are compatible with the suggested approach. It can be expected that many astronomical observations will appear to be at variance with the PCP if taken at face value. Since the tenets the suggested cosmology relies on are of the non-speculative kind (category 2a in Section 2), the reasons for such discrepancies must then be searched in possible inconsistencies, measurement errors, misinterpretations, selection effects and other missed confounding factors in the analysis of astronomical data. Another matter of interest towards which attention deserves to be turned consists in the self-regulating properties the universe must have in order for the PCP to hold.

Besides opening a range of new questions, a neo-Epicurean approach like this one also closes many questions, primarily those of cosmogonic and related kind. The phrase “the early Universe” appears in the titles of thousands of papers, but this can no longer be a topic in empirical physics. The same holds for all the unpredicted entities in the “dark sector”.

The state of physical cosmology at the beginning of the 21st century demonstrates repeatedly the undesirable lock-in effects of path dependence in science. These impose preconceptions that prevent mainstream researchers from noticing even the most obvious alternatives. Consider just the data that have led to invoking “dark energy”. An unprejudiced analysis of these suggests straightforwardly that the redshift factor $(1+z)$ is a simple exponential function of distance, but this is only told, independently of each other, by researchers who are not on the mainstream path (Ostermann, 2007, Vigoureux, Vigoureux & Vigoureux, 2008, Marosi, 2014, Traunmüller, 2014). Those on the mainstream path see a discrepancy between the empirical data and their expectations, but instead of reconsidering the traditional tenets that have given rise to these expectations, they attribute the discrepancy to the action of a phantom. In textbooks, cosmologists interpret such falsifying observations even as tantamount to the *discovery* of dark matter or dark energy. This is mentioned by Merritt (2017) after a thorough analysis of conventionalism in cosmology that corroborates my own critical view.[9]

As for the reasoning that leads from premises to conclusions, a customary *consistency check* is called for. It requires the theory to be sufficiently elaborated and errors (conceptual, logical, and mathematical) to be absent. In BB cosmology, an inconsistency has been identified in the vacant delimitation between the expanding and the non-expanding domain. The inattentive manner in which standards of measurement are treated in this context can also be understood as resulting from path dependence. It reflects the idea of the ‘rigid ruler’ of CM, which continues to be relied on despite the fact that not even space itself remains ‘rigid’ in GR.

---

[8] *Es ist das schönste Los einer physikalischen Theorie, wenn sie selbst zur Aufstellung einer umfassenden Theorie den Weg weist, in welcher sie als Grenzfall weiterlebt.*

[9] When writing previous versions of this paper, I had not yet seen Merritt (2017).

The perseverance of preconceptions that were inherited from prior theories, but which at a subsequent stage in development could be recognized as inadequate if the liberal introduction of ad hoc hypotheses was avoided, reveal a lack of awareness of the confidence problem addressed in Section 2. The *confidence check* suggested there is called for in the definition of empirical science as a pursuit of reliable statements. It requires, above all, to single out ad hoc assumptions (2c in Section 2), but it requires also to distinguish between merely *tentative* assumptions (2b) and those which appear *reliable* at the state of our knowledge (2a). Scientists often accept the tenets of established theories without reflecting about their reliability at all, while philosophers of science rather image all assumptions as fallible without distinction. None of these attitudes is suited to promote science fundamentally. Some scholars even reject the pursuit of objective observations, claiming that all observation is necessarily prejudiced, since it depends on assumptions. However, assumptions can be well-founded, and these must not be dismissed as ‘prejudices’.

The drawbacks of path dependence show themselves also in the activity of innovators. Sciama (1953) still treated GR and the expansion of the universe as givens, although the idea he investigated, inertial induction, has consequences that speak against both. Later, among the two alternatives to MoND, modified gravitation and modified inertia, only one was pursued. It was the one that can be realized by keeping CM or GR and adding some new fields to them ad hoc (Milgrom, 2002, Bekenstein, 2004) - not the one that would call the foundation of both theories into question (modified inertia). Scientific journals often publish speculative papers of the former kind, while any paper that more directly discredits the currently accepted doctrine within their field runs a very high risk of being rejected by referees and editors. Together with the similar attitude by teachers and grant providers, this leads to the perseverance of the aberrations from the path to reliable knowledge in what Kuhn (1962) called “normal science” and Lakatos (1976) “research programmes”. These labels are both sociologically appropriate, but the activities so labeled preclude any fundamental progress in science.

## Acknowledgments

Thanks are due to four anonymous reviewers of previous versions of this paper for their detailed and/or illuminative comments.

[Date of this version, 2018-02-14]

## References

Anderson, J.D., et al. (2002). Study of the anomalous acceleration of Pioneer 10 and 11. *Phys. Rev. D 65*, 082004.

Arora, S.K. (2015). *Metric contraction: an alternate model for cosmology*. Available online: https://www.academia.edu/19745143/ (accessed 2017-03-08).

Baggott, J. (2013). *Farewell to Reality: How Modern Physics has Betrayed the Search for Scientific Truth.* Pegasus Books LLC: New York.

Baryshev, Yu.V., Labini, F. S., Mantuori, M., Pietronero, L., Teerikorpi, P. (1998). On the fractal structure of galaxy distribution and its implications for cosmology. *Fractals 6*, 231–243.

Bassett, B.A., Kunz, M. (2004). Cosmic distance-duality as a probe of exotic physics and acceleration. *Phys. Rev. D 69,* 101305(R).

Bekenstein, J.D. (2004). Relativistic gravitation theory for the modified Newtonian dynamics paradigm. *Phys. Rev. D 70,* 083509.

Blondin, S., et al. (2008). Time dilation in type Ia supernova spectra at high redshift. *Astrophys. J. 682*, 724–736.

Bondi, H., Gold, T. (1948). The steady-state theory of the expanding universe. *Mon. Not. R. Astron. Soc. 108,* 252–270.

Bouwens, R.J., Illingworth, G.D., Blakeslee, J.P., Broadhurst, T.J., Franx, M. (2004). Galaxy size evolution at high redshift and surface brightness selection effects: constraints from the Hubble Ultra Deep Field. *Astrophys. J. 611*, L1–L4. DOI 10.1086/423786

Caldwell, R.R., Dave, R., Steinhardt, P.J. (1998). Cosmological Imprint of an Energy Component with General Equation-of-State. *Phys. Rev. Lett. 80* (8): 1582–1585.

David, P.A. (1985). Clio and the Economics of QWERTY. *Am. Econ. Rev. 75* (2), 332–337.

de Sitter, W. (1917). Einstein's theory of gravitation and its astronomical consequences. Third paper. *Mon. Not. R. Astron. Soc. 78*, 3–28.

Disney, M.J., et al. (2008). Galaxies appear simpler than expected. *Nature 455*, 1082–1084.

Djorgovski, S., Spinrad, H. (1981). Toward the application of metric size function in galactic evolution and cosmology. *Astrophys. J. 251*, 417–422.

Edwards, M.R. (1998). Evidence of a cosmological matter and energy cycle. *Apeiron 5*, 157–163.

Edwards, M.R. (2012). Does the Hubble redshift flip photons and gravitons? *Astrophys. Space Sci. 339*, 13–17.

Einstein, A. (1907). Über das Relativitätsprinzip und die aus demselben gezogenen Folgerungen, Teil V. Relativitätsprinzip und Gravitation. *Jahrbuch der Radioaktivität und Elektronik 4*, 454–462.

Einstein, A. (1911). Über den Einfluß der Schwerkraft auf die Ausbreitung des Lichtes. *Ann. Phys. 35*, 898–908.

Einstein, A. (1912). Gibt es eine Gravitationswirkung, die der elektrodynamischen Induktionswirkung analog ist? *Vierteljahrschrift für gerichtliche Medizin und öffentliches Sanitätswesen, 44*, 37–40.

Einstein, A. (1917a ). Kosmologische Betrachtungen zur allgemeinen Relativitätstheorie. *Sitzber. K. Preuss. Aka. Wiss. VI*, 142–152.

Einstein, A. (1917b). *Über die spezielle und die allgemeine Relativitätstheorie*. Vieweg: Braunschweig.

Etherington, I.M.H. (1933). On the definition of distance in general relativity. *Phil. Mag. 15*, 761–773.

Famaey, B., Gentile, G., Bruneton, J-P., Zhao, H.S. (2007). Insight into the baryon-gravity relation in galaxies. *Phys. Rev. D 75*, 063002.

Famaey, B., McGaugh, S. (2012). Modified Newtonian Dynamics: Observational Phenomenology and Relativistic Extensions. *Living Rev. Relativity 15*, 10 (164p).

Farrar, G.R., Rosen, R.A. (2007). A new force in the dark sector? *Phys. Rev. Lett. 98,* 171302.

Filippenko, A.V., Riess A.G. (1998). Results from the High-z Supernova Search team. *Phys Rep. 307*, 31–44.

Foley, R.J., et al. (2005). A definitive measurement of time dilation in the spectral evolution of the moderate-redshift type Ia supernova 1997ex. *Astrophys. J. 626,* L11–L14.

Fukugita, M., Peebles, P.J.E., (2004). The cosmic energy inventory, Astrophys. J. 616, 643–668.

Gettier, E.L. (1963). Is justified true belief knowledge?, *Analysis 23*, 121-123.

Giulini, D. (2014). Does cosmological expansion affect local physics? *Studies in History and Philosophy of Modern Physics 46*, 24–37.

Goldhaber, G., et al. (1997). Observation of cosmological time dilation using Type Ia supernovae as clocks. In: *Thermonuclear Supernovae*, Ruiz-Lapuente P., Canal R., Isern J., Eds.; Kluwer: Dordrecht, pp. 777–784.

Goldhaber, G., et al. (2001). Timescale stretch parameterization of type Ia supernova B-band light curves. *Astrophys. J. 558*, 359–368.

Guth, A.H. (1981). The inflationary universe: a possible solution to the horizon and flatness problems. *Phys. Rev. D 23,* 347–356.

Guth, A.H., Kaiser, D.I., Nomura, Y. (2014). Inflationary paradigm after Planck 2013, *Phys. Lett. B. 733*, 112–119.

Harwit, M. (2006). *Astrophysical Concepts*. 4th ed. Springer Science & Business Media, pp. 572–576.

Hoefer, C. (1994) Einstein’s struggle for a Machian graviatation theory. *Studies in History and Philosophy of Science Part A, 25*, 287-335.

Holanda, R.F.L., Lima, J.A.S., Ribeiro, M.B. (2011). Cosmic distance duality relation and the shape of galaxy clusters. *Astron. Astrophys. 528,* L14. DOI 10.1051/0004-6361/201015547

Holwerda, B.W., et al. (2014). The sizes of candidate z ~ 9–10 galaxies; confirmation of the bright CANDLES sample and correlation with luminosity and mass. *Astrophys. J. 808*, 6. DOI 10.1088/0004-637X/808/1/6

Ijjas, A., Steinhardt, P.J., Loeb, A. (2014). Inflationary schism. *Phys. Lett. B. 736*, 142–146.

Iorio, L. (2011). An empirical explanation of the anomalous increases in the astronomical unit and the lunar eccentricity. *Astron. J. 142*, 68 (3p).

Jolink, A., Vromen, J. (2001). Path dependence in scientific evolution. In *Evolution and Path Dependence in Economic Ideas*; Garrouste, P., Ioannides, S. Eds.; Elgar: Cheltenham, pp. 205–224.

Kashlinsky, A., Atrio-Barandela, F., Kocevski, D., Ebeling, H. (2008). A measurement of large-scale peculiar velocities of clusters of galaxies: results and cosmological implications. *Astrophys. J. Lett. 686*, L49–L52. DOI 10.1086/592947

Krasinsky, G.A., Brumberg, V.A. (2004). Secular increase of astronomical unit from analysis of the major planets motions, and its interpretation. *Celest. Mech. & Dyn. Astron. 90*, 267–288.

Křížek, M., Somer, L. (2013). Antigravity – its manifestations and origin. *Int. J. Astron. Astrophys. 3*, 227–235.

Kuhn, T. (1962). *The Structure of Scientific Revolutions*. Univ. Chicago Press: Chicago.

Lahav, O., Massimi, M. (2014). Dark energy, paradigm shifts, and the role of evidence. *Astron. Geophys. 55*, 3.12–3.15.

Lakatos, I. (1976). Falsification and the methodology of scientific research programmes. In: *Criticism and the Growth of Knowledge*, Lakatos I., Musgrave A., Eds.; Cambridge University Press: London, pp. 170–196.

Leibundgut, B., et al. (1996). Time dilation in the light curve of the distant Type IA supernova SN1995K. *Astrophys. J. Lett. 466*, L21.

Lemaître, G. (1927). Un Univers homogène de masse constante et de rayon croissant rendant compte de la vitesse radiale des nébuleuses extra-galactiques. *Ann. Soc. Sci. Brux. 47*, 49–59.

López-Corredoira, M. (2010). Angular size test on the expansion of the universe. *Int. J. Mod. Phys. D 19*, 245–291.

Mach, E. (1889) *Die Mechanik in ihrer Entwicklung historisch-kritisch dargestellt*. 2nd ed., Brockhaus: Leipzig.

Marosi, L.A. (2014). Hubble diagram test of 280 supernovae redshift data. *J. Modern Physics 5*, 29–33.

Masreliez, C.J. (2004). Scale Expanding Cosmos Theory I – An introduction. *Apeiron 11*, 99–133.

Merritt, D. (2017). Cosmology and convention, *Studies in Hisotry and Philosophy of Modern Physics, 57*, 41-52.

Milgrom, M. (1983). A modification of the Newtonian dynamics as a possible alternative to the hidden mass hypothesis. *Astrophys. J. 270*, 365–370.

Milgrom, M. (2002). MOND – theoretical aspects. *New Astron. Rev. 46*, 741–753.

Milgrom, M., Sanders, R.H. (2007). MOND rotation curves of very low mass spiral galaxies. *Astrophys. J. 658*, L17–L20.

Moffat, J.W. (2005). Gravitational theory, galaxy rotation curves and cosmology without dark matter. *J. Cosmol. Astropart. P. 05*. DOI 10.1088/1475-7516/2005/05/003

Mosleh, M., et al. (2012). The evolution of mass-size relation for Lyman break galaxies from z = 1 to z = 7. *Astrophys. J. Lett. 756*, L12.

Mößbauer, R.L. (1998). History of neutrino physics: Pauli's letters. In *Proceedings of the Fourth SFB-375 Ringberg Workshop "Neutrino Astrophysics"*, 3–5, https://arxiv.org/pdf/astro-ph/9801320.pdf#page=11 (accessed 2017-03-09).

Ostermann, P. (2007). Das relativistische Modell eines stationären Hintergrunduniversums und die Supernova-Ia-Daten. Conference contribution, *DPG 07*, GR-205.2, http://www.peter-ostermann.de/assets/07a.pdf (accessed 2017-03-09).

Peacock, M.S. (2009). Path dependene in the production of scientific knowledge. *Social Epistemology 23*, 105–124.

Peebles, P.E.J., Ratra, B. (2003). The cosmological constant and dark energy. *Rev. Mod. Phys. 75*, 559–606.

Perlmutter, S., et al. (1998). Discovery of a supernova explosion at half the age of the Universe. *Nature 391*, 51–54.

Pitjeva, E.V., Pitjev, N.P. (2012). Changes in the Sun's mass and gravitational constant estimated using modern observations of planets and spacecraft. *Solar Syst. Res. 46*, 78–87.

Popper, K. (1935). *Logik der Forschung: zur Erkenntnistheorie der modernen Naturwissenschaft*. Julius Springer: Wien, [The Logic of Scientific Discovery].

Riess, A.G., et al. (1997). Time dilation from spectral feature age Measurements of Type Ia Supernovae. *Astron. J. 114*, 722–729.

Riess, A.G., et al. (1998). Observational evidence from supernovae for an accelerating universe and a cosmological constant. *Astron. J. 116*: 1009–1038.

Rubin, V., Ford, W.K. Jr., Thonnard, N. (1980). Rotational properties of 21 Sc galaxies with a large range of luminosities and radii, from NGC 4605 (R = 4 kpc) to UGC 2885 (R = 122 kpc). *Astrophys. J. 238*, 471–487.

Sandage, A. (1987). Observational cosmology 1920–1985. In *Observational cosmology; Proceedings of the IAU Symposium, Beijing, People's Republic of China, Aug. 25–30, 1986* (A88-29629 11–90). D. Reidel Publishing Co.: Dordrecht, pp. 1–27. (http://adsabs.harvard.edu/abs/1987IAUS..124....1S)

Sandage, A.R. (1972). The redshift–distance relation. I. Angular diameter of first ranked cluster galaxies as a function of redshift: the aperture correction to magnitudes. *Astrophys. J. 173*, 485–498.

Sanders, R.H. (2003). Clusters of galaxies with modified Newtonian dynamics. *Mon. Not. R. Astron. Soc. 342*, 901–908.

Sciama, D.W. (1953). On the origin of inertia. *Mon. Not. R. Astron. Soc. 113*, 34–42.

Shibuya, T., Ouchi, M., Harikane, Y. (2015). Morphologies of ~ 190,000 galaxies at z=0–10 revealed with HST legacy data. I. Size evolution. *Astrophys. J. Suppl. 219*, 15.

Swaters, R.A., Sanders, R.H., McGaugh, S.S. (2010). Testing modified Newtonian dynamics with rotation curves of dwarf and low surface brightness galaxies. *Astrophys. J. 718*, 380–391.

Tolman, R.C. (1930). On the estimation of distances in a curved universe with a non-static line element. *Proc. Natl. Acad. Sci. USA 16*, 511–520.

Traunmüller, H. (2014). From magnitudes and redshifts of supernovae, their light-curves, and angular sizes of galaxies to a tenable cosmology. *Astrophys. Space Sci. 350*, 755–767.

Turyshev, S.G, Toth, V.T. (2010). The Pioneer anomaly. *Living Rev. Relativ.*, *13*, 4–175.

Turyshev, S.G., et al. (2012). Support for the thermal origin of the Pioneer anomaly. *Phys. Rev. Lett. 108*, 241101.

Unzicker, A., Schmidle, D. (2007). *A quick and dirty approach to verify the Pioneer anomaly*. https://arxiv.org/pdf/gr-qc/0702151.pdf (accessed 2017-03-09).

van der Wel, A., et al. (2008). Recent structural evolution of early-type galaxies: size growth from z = 1 to z = 0. *Astrophys. J. 688*, 48–58.

Vigoureux, J.-M., Vigoureux, P., Vigoureux, B. (2008). Cosmological applications of a geometrical interpretation of “c”. *Int. J. Theor. Phys. 47*, 928–935.

Watkins, R., Feldman, H.A., Hudson, M.J. (2009). Consistently large cosmic flows on scales of 100 h−1 Mpc: a challenge for the standard ΛCDM cosmology. *Mon. Not. R. Astron. Soc. 392*, 743–756.

Wei, H. (2010). Observational constraints on cosmological models with the updated long gamma-ray bursts. *J. Cosmol. Astropart. P. 2010*. DOI 10.1088/1475-7516/2010/08/020

Yang, X., Yu, H.-R., Zhang, Z.-S., Zhang, T.-J. (2013). An improved method to test the distance-duality relation. *Astrophys. J. 777*, L24.

Zwicky, F. (1933). Die Rotverschiebung von extragalaktischen Nebeln. *Helv. Phys. Acta 6*, 110–127.

Zwicky, F. (1937). On the masses of nebulae and of clusters of nebulae. *Astrophys. J. 86*, 217–246.